# Multipixel characterization of imaging CZT detectors for hard X-ray imaging and spectroscopy

S. V. Vadawale[a*], J. Hong[a], J. Grindlay[a], P. Williams[a], M. Zhang[a], E. Bellm[a],
T. Narita[b], W. Craig[c], B. Parker[d], C. Stahle[d], Feng Yan[d]

[a]Harvard-Smithsonian Center for Astrophysics, Cambridge, MA, USA
[b]College of the Holly Cross , Worcester, MA, USA
[c]Lawrence Livermore National Laboratory, Livermore, CA, USA
[d]NASA/Godard Space Flight Center, Greenbelt, MD, USA

## ABSTRACT

We report our in-depth study of Cd-Zn-Te (CZT) crystals to determine an optimum pixel and guard band configuration for Hard X-ray imaging and spectroscopy. We tested 20x20x5mm crystals with 8x8 pixels on a 2.46mm pitch. We have studied different types of cathode / anode contacts and different pixel pad sizes. We present the measurements of leakage current as well as spectral response for each pixel. Our I-V measurement setup is custom designed to allow automated measurements of the I-V curves sequentially for all 64 pixels, whereas the radiation properties measurement setup allows for interchangeable crystals with the same XAIM3.2 ASIC readout from IDEAS. We have tested multiple crystals of each type, and each crystal in different positions to measure the variation between individual crystals and variation among the ASIC channels. We also compare the same crystals with and without a grounded guard band deposited on the crystal side walls vs. a floating guard band and compare results to simulations. This study was carried out to find the optimum CZT crystal configuration for prototype detectors for the proposed Black-Hole Finder mission, EXIST.

**Keywords:** CZT characterization, multipixel measurements, hard X-ray imaging,

## 1. INTRODUCTION

Over the past several years, Cadmium-Zinc-Telluride (CZT) has emerged as the material of choice for hard X-ray / soft gamma-ray detectors. CZT offers many advantages over the conventional hard X-ray detectors such as scintillator or Si/Ge based semi-conductor detectors. Due to its relatively large band-gap, CZT can be operated at room temperature avoiding very expensive and difficult cryogenic cooling required for the Si/Ge semi-conductors. However, even with the larger band-gap, the total energy required for generating one electron-hole pair in CZT is much less then that required for scintillators, resulting in much better energy resolution. The high effective atomic number of CZT results in high detection efficiency up to 600 keV for reasonable thickness. Another big advantage of CZT is the possibility of imaging by subdividing anode into multiple pixels. Pixelated crystals also show improved spectral performance due to the *small pixel* effect[1]. Because of these obvious advantages, CZT has been selected as hard X-ray detector in many future X-ray missions such as Swift[2,] Constellation-X[3] and EXIST.

For the past few years, our group at Harvard has been developing CZT based hard X-ray detectors mainly for the proposed Black-hole finder probe, EXIST[4]. According to the present baseline design, EXIST will have coded aperture mask telescopes having a tiled CZT array as a detector plane. The prime scientific motivation for EXIST is to conduct an all sky hard X-ray (10 - 600 keV) survey with the sensitivity comparable to the ROSAT all sky survey in the soft X-rays (0.1 - 2 keV), which necessitates a very large area of about 8 m$^2$ for the detector plane. Another important scientific requirement is, imaging resolution of 5 arcmin, which requires the individual pixel, on the detector plane to have 1.25 mm pitch for the 1.5 m mask-detector spacing in the baseline design. Thus EXIST will require about 5 million pixels. Given the fact that the CZT crystal yield sharply decreases with increasing size, a crystal size of 20 x 20 mm seems to be an optimum size for EXIST and thus each CZT crystal will have at least 256 pixels.

---

* Send correspondence to S. V. Vadawale(svadawale@cfa..harvard.edu)

So far there have been many studies of basic properties of CZT as an X-ray detector, however most of these studies are done with either planar detectors or for relatively few pixels in case of pixelated detectors. In real application, however, all the pixels of all CZT crystals are going to be used for the sky imaging. Therefore it is imperative to carry out the study of basic properties for all pixels of a given CZT crystal. This led us to our present study, where the main aim is to identify the optimum crystal configuration, in terms of CZT material, metallic contacts, pad sizes, guard band etc., based on the all pixel characterization of each crystal. Specific goals of our study are to:
- Characterize the performance of individual pixels
- Identify any non-uniformity between various pixels of the same crystal
- Study consistency of performance across multiple crystals of same type
- Study the effects of various guard-band configurations
- Search for optimal metallic contact combination

## 2. EXPERIMENTAL SETUP

### 2.1 CZT Crystals

We used CZT crystals manufactured by IMARAD Imaging Systems, Israel for our study. IMARAD CZT crystals are grown by the modified Horizontal Bridgman method[5]. It is a known fact that IMARAD CZT material is slightly N-type material[6]. The crystal properties such as resistivity depend on the metallic contact. The low work-function metals such as Indium forms an *ohmic contact* with IMARAD CZT crystals whereas the high work-function metals such as Gold and Platinum forms a high resistance contact, or *blocking contact*[7]. Typical IMARAD CZT crystal comes in 19.44 x 19.44 x 5 mm size and with Indium contacts on both anode and cathode surface. The anode surface is pixelated in 8 x 8 array of pixels with pitch of 2.46 mm. Typical gap between pixel pads is 0.6 mm.

We obtained total 20 crystals, 9 with standard Indium contacts and 11 without any metallic contact, for our testing. We characterized performance of multiple crystals with Indium/Indium as well as Platinum/Platinum contacts on cathode/anode. We also tested few crystals with hybrid contacts i.e. Platinum on cathode and Indium or Aluminum on anode. The metalization of all crystals except standard IMARAD (Indium/Indium) crystals was carried out at NASA/Goddard and at College of Holycross using standard evaporation technique.

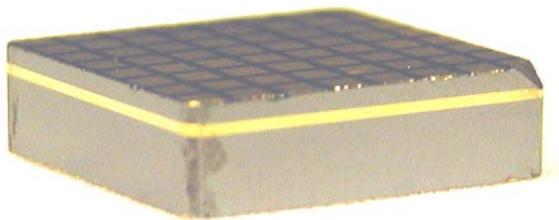
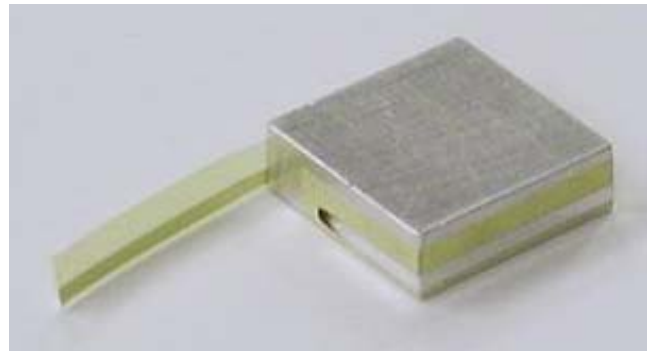

**Figure 1.** Two types of guard-bands. Picture on left shows our concept of the grounded guard-band, where the Gold strip is deposited on the side walls of the crystal. It is connected to a corner pixel through a beveled corner. This pixel is then connected to ground. The image on right shows IMARAD type floating guard-band. It is an adhesive tape with a metallic strip deposited near one edge, which is wrapped around the crystal (crystal shown in the picture is a dummy Aluminum crystal).

We tested these crystals with and without IMARAD style guard-band. As a concept study, we also carried out testing of grounded guard-band on the crystal wall with Gold/Gold contacts. It is important to understand these two types of guard-bands in detail (see Figure 1). Standard CZT crystals from IMARAD come with a special band wrapped around the crystal walls. This band is basically a plastic adhesive tape with metal deposition near one side. This 5 mm (same as crystal thickness) wide tape is simply wrapped on the walls of the crystals. According to IMARAD, this guard-band significantly improves the performance of pixels on the edge of the crystals. We have tested this and report results below. The other type of guard-band we tried, a ground wall guard-band, is a metal strip deposited on the crystal walls close to the anode surface and physically connected to ground. The motivation to test this concept was to reduce the leakage current on the edge pixels. It is well known that due to the current through the crystal walls, effective leakage current on the edge pixels is generally higher then the inner pixels. The classical approach to overcome the large wall current is to put a grounded guard-ring around the outermost anode surface. However, in the tiled array of pixelated crystals, such a guard-ring is not suitable because of two reasons, it reduces the effective detector area and it becomes difficult to maintain pixel pitch across multiple crystals. This led us to consider the idea of putting the classical guard-ring on the walls. We tested this concept with Gold contacts. We deposit Gold on cathode/anode surface as well as in a small (~1 mm) strip near anode surface. In order to ground the ring on the walls, we bevel one of the corners (see Figure 1) so as to connect the corner pixel pad with the guard band on the wall. One limitation of this scheme that one corner pixel has to be sacrificed in order to ground the guard-band. We tried out this scheme for few crystals with Gold/Gold contacts. For each crystal, we measured leakage and radiation data (total counts, photo peak counts, energy resolution etc.) for each pixel. We also repeated these measurements after rotating the crystal position by 180 degree, to compensate for any systematic bias in the measurement electronics for various pixels.

**2.2 Leakage Current Measurement**

We have designed a custom setup which can automatically measure leakage current for each of the 64 pixels of the 8x8 crystal for any bias voltage ranging from 0 to -1000 V. The schematic diagram of our setup is shown in Figure 2. The custom designed 'selector board' is the most important component of the system, which performs the actual pixel selection. It consists of 64 SPDT (Single Pole Double Through) switches and is controlled by PC over the parallel port. At any time, any one of the 64 channels can be selected for the current measurement while the rest of the 63 channels are connected to ground. The current measurement is performed by the *Keithley* electrometer (model 237), which is controlled by PC over a GPIB interface. The Keithley also generates and controls the low voltage input necessary to control the HV supply (NIM module Canberra 302).

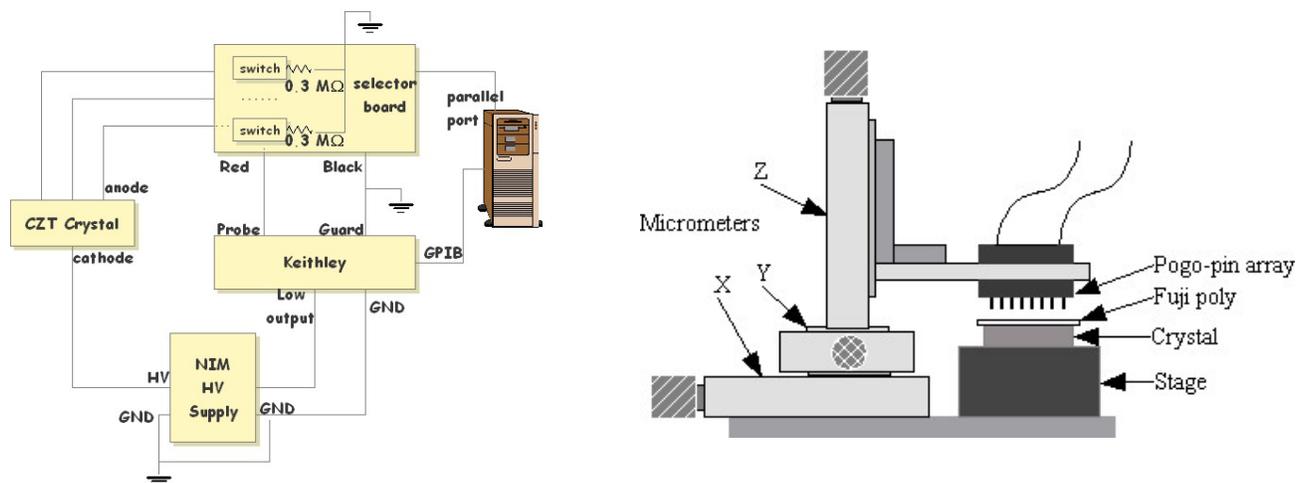

**Figure 2:** Schematic diagram of our leakage current measurement setup (left) and mechanical scheme for mounting the crystals (right).

The mechanical setup for mounting the crystal consists of an 8x8 array of spring loaded pogo-pins (see Figure 2). The pogo pin array is attached to a movable stage. Three micrometers control its motion in all three directions. The crystal is placed on a fixed stage with the cathode surface facing down. The stage also provides the electrical connection to apply the necessary bias voltage to the crystal. The pogo-pins connect to the anode pixels through a vertically conductive rubber pad (Fujipoly). The micrometer controlled motion and spring action of the pogo-pins provides sufficient pressure to connect through Fujipoly, at the same time ensures that there is no permanent damage to either Fujipoly or the crystal surface.

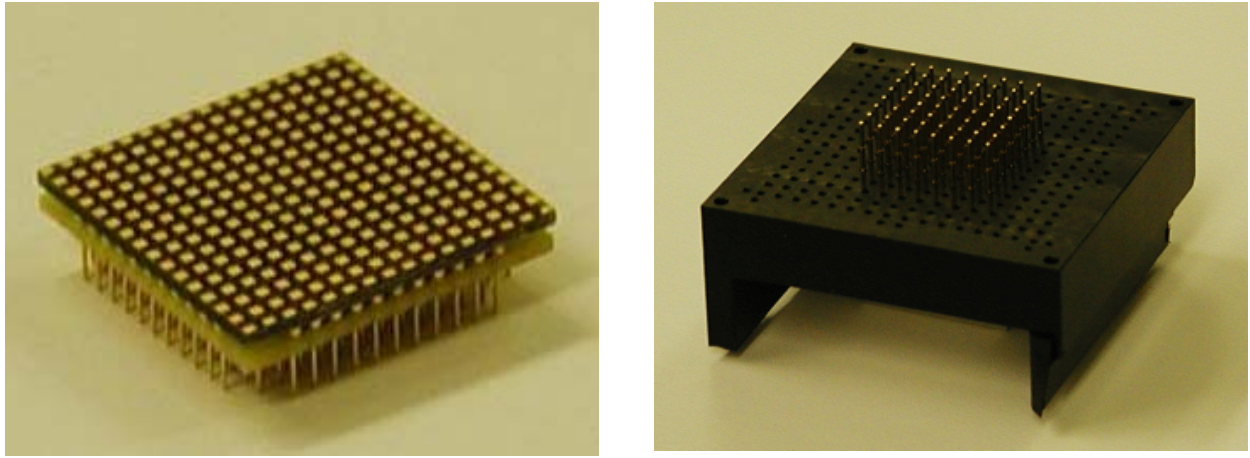

**Figure 3:** Bare COB (left) and pogo-pin array system (right). The bare COB contains two XAIM 3.2 ASICs reading out the full 16x16 pixel array. The pogo-pin array or "nail-bed" system has holes precision-machined in a Delrin block and matching with the bare COB input pads. Spring loaded pogo-pins inserted in these holes provide electrical connection with the ASIC input.

In a typical measurement, the PC first sets the appropriate bias voltage through the Keithley. Then it selects one pixel through the selector board and reads the current measured by the Keithley. This process is repeated for a given range of pixels and given range of bias voltage. We measure the leakage current from all 64 pixels for a bias voltage range of 0 to -1000 V in 20 steps. We wait for about 500 ms after selecting any pixel so as to stabilize the current measurement. Similarly we provide about 10 s delay between two bias voltage settings to stabilize the bias voltage at the desired level.

**2.3 Radiation properties measurements**

Our radiation properties measurement setup is based on the XAIM 3.2 ASIC (Application Specific Integrated Circuit) from IDEAS Corporation. The XAIM is a 128 channel, self-triggered and data driven ASIC[8]. It features an externally adjustable threshold. When any of the 128 input channels exceeds the specified threshold, the ASIC generates a trigger and the amplified signal and the channel number are available on the ASIC output for a fixed (adjustable) period of time. Typically XAIM ASICs are available as a complete module from IMARAD, called a COB (Crystal On Board) consisting of a 2 x 2 array (4 x 4 cm) of IMARAD CZT crystals and two XAIM ASICs, giving a 16 x 16 pixel array detector. We specifically obtained five bare COBs (without CZT crystal bonded with ASIC) to carry out tests of multiple CZT crystals with the same ASIC (Figure 3). We also obtained the XA Controller Readout system from IDEAS to control and readout from these ASICs. The XA Controller Readout system consists of a PLD-based XA Controller and a motherboard capable of housing 2 x 2 COBs (total 8 x 8 cm or 32 x 32 pixel CZT array). The motherboard, XA Controller and PC are connected by 68 pin mini SCSI cable. The high voltage bias is applied using an external NIM module. We have previously reported early imaging results from this system in Narita et al. (2002)[9].

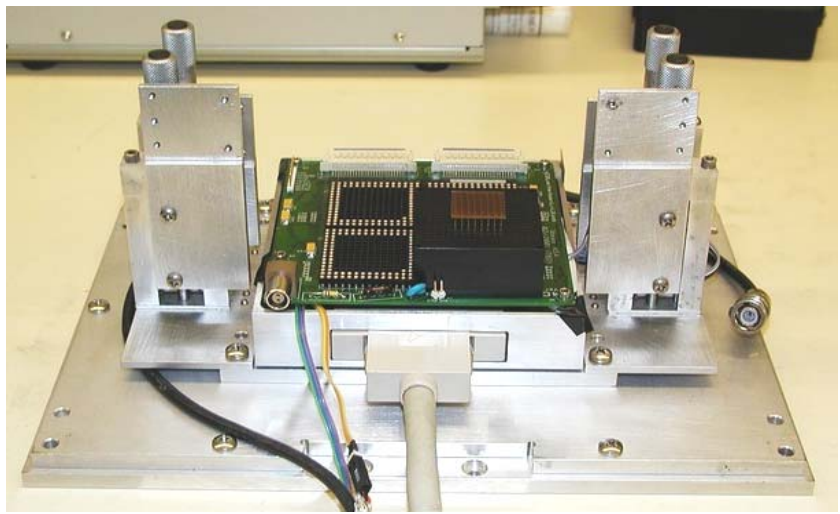

**Figure 4:** The nail-bed system mounted on the motherboard of the XA Controller system. The crystal is kept on the Fujipoly placed on top of the pogo-pins. The HV bias board is mounted on the micrometer controlled studs on two sides.

The mechanical setup for mounting the CZT crystal consists of a nail-bed and a high voltage bias board. The nail-bed system is made from a Delrin block and has small holes matching with the connection pad on the bare COB (see Figure 4). The crystal and Fujipoly are mounted on the spring loaded pogo-pins inserted in these holes. The high voltage bias board is mounted on micrometer controlled studs and provides the required pressure to connect through the Fujipoly. The micrometer-controlled motion and the spring action of pogo-pins ensure that there is no damage to either crystal surface or Fujipoly due to excess pressure. The entire assembly is kept in a light tight box and the radiation source is placed at a distance of about 30 cm. Before saving data we ensure, by looking at the image of crystal, that the crystal is illuminated uniformly by X-rays. We save data consisting of energy and pixel number for each individual X-ray event. For appropriate energy calibration of each pixel and corresponding ASIC channel, we take data from two radiation sources, $Am^{241}$ (giving a line at 60 keV) and $Co^{57}$ (giving a line at 122 keV). The low energy threshold of our system is approximately 30 keV. We fit the high energy edge of the X-ray line with a Gaussian to find the FWHM in ADC channels and then convert it to absolute energy using two point energy calibration of the individual ASIC channel.

## 3. RESULTS

We have measured leakage current and radiation properties for large number of CZT crystals of different types (see Table 1). A typical result of leakage current measurement is shown in Figure 5. The top two plots show variation of leakage current (left plot) and corresponding variation in resistivity of the crystal (right plot) with bias voltage. The bottom left plot shows the leakage current for each of 64 pixels at -960 V. The image on the bottom right corner shows the leakage current map of the crystal. The results of radiation measurements for the same crystal are shown in Figures 6a and 6b. We generate similar plots for the leakage current as well as radiation properties measurement for all crystals tested. Thus we have measurements of six quantities for each of the 64 pixels for any crystal, giving a very large database of these results. Therefore, in order to study the systematic trends in the pixels of the same crystal and to have meaningful comparison between the crystals of the same type as well as crystals of different types, we take the average values of these quantities for the pixels with similar locations within a crystal. Specifically, we take the average of the inner 16 pixels (4 x 4 array), next ring of 20 pixels, and 28 pixels on the edges. The variation of the two most important quantities, leakage current and energy resolution, across the crystals of various types is shown in the Figures 7a-d. In the following sub-sections we summarize the results of these measurements for crystals with different types of metallic contacts on cathode/anode.

**Table 1:** Results of our tests of various metallic contacts with IMARAD CZT crystals

| Sr. No. | Contact type cathode/anode | No. of crystal | Average[a] leakage current | | | Average[a] energy resolution | | |
|---|---|---|---|---|---|---|---|---|
| | | | Inner 16 | Middle 20 | Edge 28 | Inner 16 | Middle 20 | Edge 28 |
| 1 | In/In with fgb[b] | 9 | 18.4 | 19.3 | 18.5 | 6.5 | 6.8 | 7.2 |
| 2 | In/In without fgb | 2 | 12.8 | 13.8 | 14.5 | 6.6 | 6.7 | 7.8 |
| 3 | Cry. in 2 with fgb | 2 | 13.5 | 14.5 | 14.4 | 6.4 | 6.6 | 7.0 |
| 4 | Pt/Pt with fgb | 8 | 0.4 | 0.5 | 2.2 | 5.2 | 5.5 | 6.2 |
| 5 | Pt/Pt without fgb | 3 | 0.5 | 0.9 | 10.5 | 5.6 | 6.2 | 8.9 |
| 6 | Cry. in 5 with fgb | 3 | 0.4 | 0.5 | 1.8 | 5.1 | 5.7 | 6.5 |
| 7 | Au/Au with fgb | 1 | 0.1 | 0.1 | 7.4 | 8.1 | 8.0 | 9.3 |
| 8 | Au/Au without fgb | 1 | 1.3 | 1.8 | 9.4 | 7.4 | 8.2 | 13.8 |
| 9 | Au/Au with half ggb | 1 | 1.3 | 1.6 | 9.3 | 9.0 | 10.3 | 78.9 |
| 10 | Au/Au with full ggb | 1 | 6.2 | 5.7 | 3.2 | 5.7 | 5.9 | 7.0 |
| 11 | Pt/Al | 1 | 3.2 | 8.3 | 13.2 | 6.7 | 7.8 | 8.5 |
| 12 | Pt/In | 2 | 2.8 | 3.3 | 2.8 | 5.7 | 6.4 | 8.3 |

**a:** For the spread in these quantities among individual crystals see Figures 7a-d
**b:** fgb = floating guard-band, ggb = grounded guard-band

### 3.1 Indium cathode - Indium anode

Results of our measurements for IMARAD CZT crystals with Indium/Indium contacts are summarized in Figure 7a. We find that the Indium contacts with IMARAD CZT exhibit ohmic contact behavior, as we have previously reported[9]. The leakage current across all pixels increases linearly with bias voltage. We find that there is no systematic variation in the leakage current between inner and edge pixels of a given crystal, for the standard IMARAD crystals (which come with floating guard-band). However, we find that the average values of leakage current for different crystals ranges from 10 nA to 30 nA among the sample of nine crystals we tested. Radiation properties of IMARAD CZT crystals with In/In contacts show mixed behavior across various pixels of individual crystals. The total count rate and the photo-peak efficiency (defined as the ratio of counts in the photo-peak energy ± FWHM to total counts for each pixel) are similar for all pixels within the statistical fluctuations. The energy resolution for the In/In crystals ranges between 6 - 8 % at 122 keV. However, the edge pixels show poor energy resolution (by about 0.5 %) compared to the inner pixels of the same crystal. For the two crystals we tested with and without the floating guard-band,

### 3.2 Platinum cathode - Platinum anode

Results of our measurements for IMARAD CZT crystals with Platinum/Platinum contacts are summarized in Figure 7b. IMARAD crystals with Platinum contacts show systematic variation across various pixels. Platinum contacts exhibit blocking contact behavior in which the leakage current saturates after a particular bias voltage, however the saturation bias varies among inner, middle and edge pixels. The edge pixels show 1 - 10 nA leakage current which is an order of magnitude higher then inner pixels which show 0.1 - 1 nA. Radiation properties show similar trend among the edge and inner pixels. We get about 5 % energy resolution for inner pixels whereas edge pixels show resolution degraded by about 1 - 2 %. Overall we find that the Platinum contacts give better performance then Indium contacts in terms of both leakage current as well as energy resolution for inner pixels.

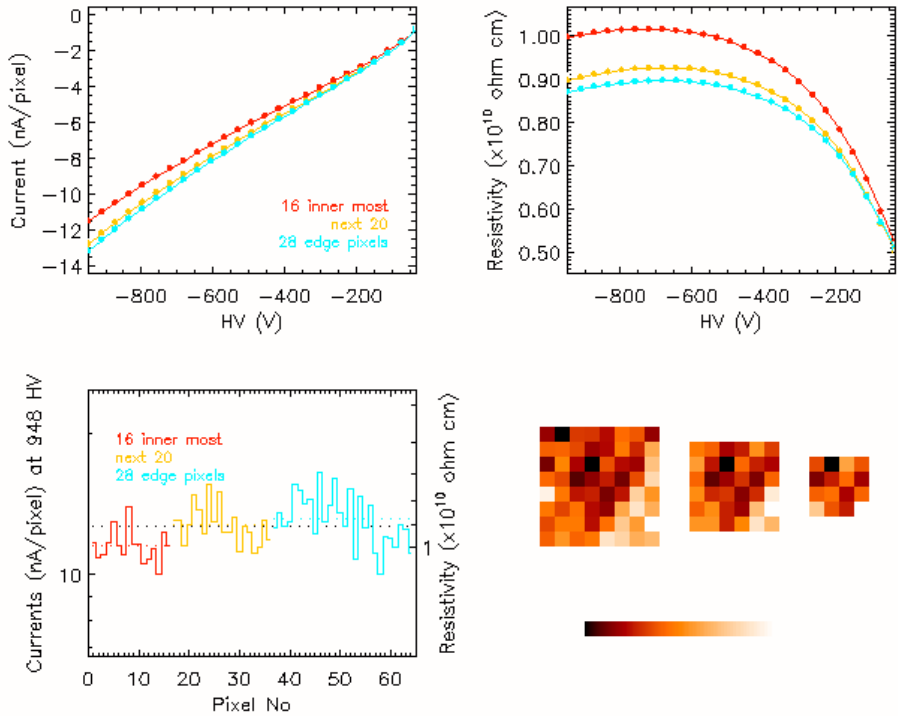

**Figure 5:** Representative plots summarizing our leakage current measurements for each crystal. The top two plots show the average leakage current (left) and resistivity (right) of the inner 16, middle 20 and edge 28 pixels vs. bias voltage. The bottom left plot shows the leakage current of each pixel at -950 V. The map on the bottom right shows the distribution of the leakage current across all pixels. We generate similar plots for all crystals.

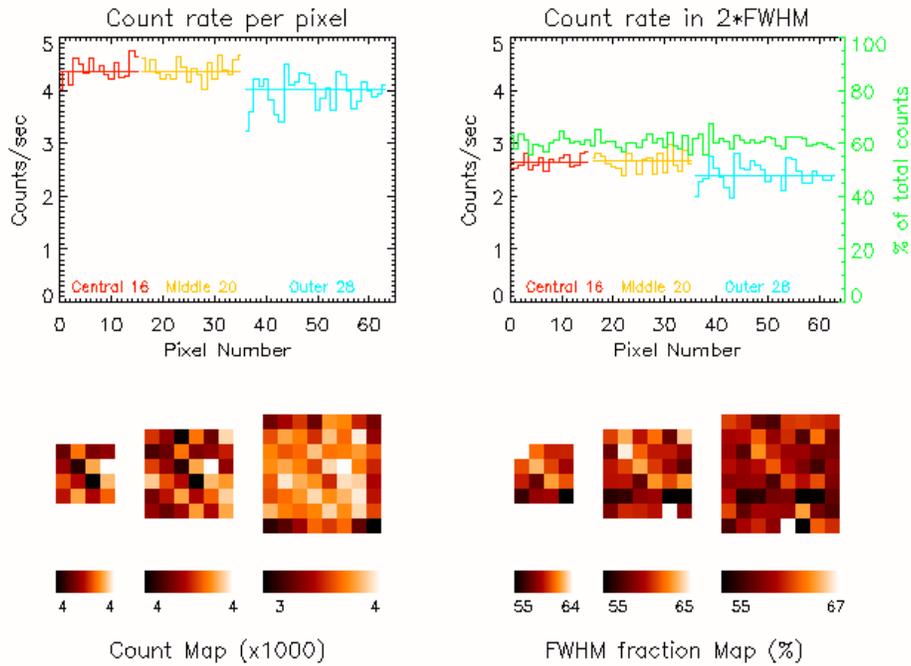

**Figure 6a:** Representative plots showing the total counts in each pixel (left) and within the photopeak ± FWHM (right) for each pixel of an Indium/Indium crystal. Maps at the bottom show distribution of total counts and fraction of total counts within the photopeak.

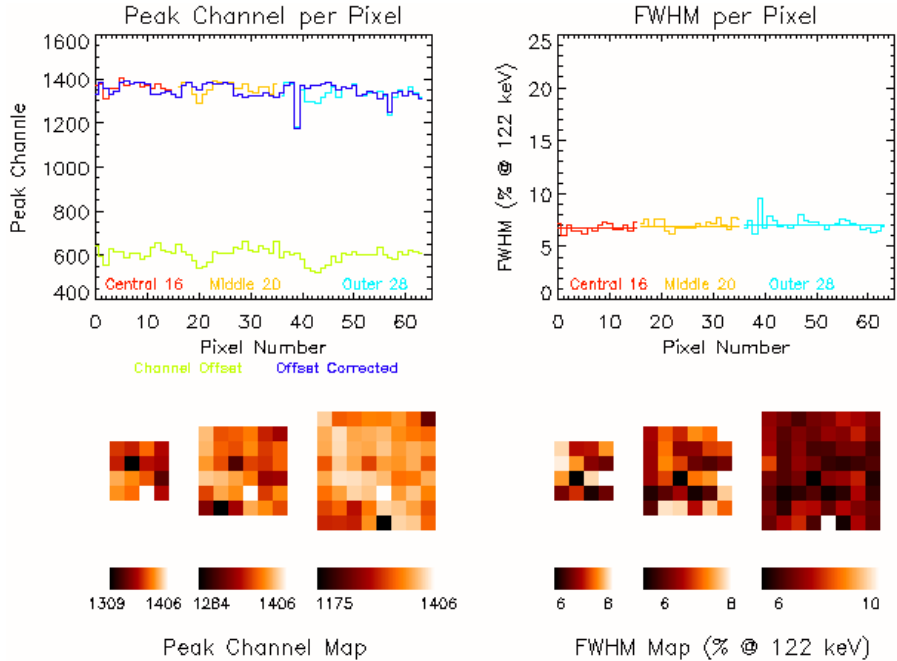

**Figure 6b:** Representative plot showing peak channel (indicating gain of each ASIC channel) and energy resolution per pixel of an Indium/Indium crystal. Also shown in the top left plot are the offset values of each ASIC channel and offset corrected peak channel. Maps at the bottom show distribution of these quantities within the crystal. We generate similar maps for every crystal.

### 3.3 Effect of the floating guard-band

We tested two Indium/Indium and three Platinum/Platinum crystals with and without IMARAD type floating guard-band. We find that this guard-band does not have significant effect on the leakage current of the inner vs. outer pixels of the In/In crystals. In both the crystals we tested without floating guard-band, the average leakage current as well as the statistical variation among different pixels is similar to what we get for the same crystals with floating guard-band. However, for the same In/In crystals the energy resolution of the edge pixels is improved from about 8 % to about 6.5 % with the floating guard-band. For the Pt/Pt crystals, on the other hand, the floating guard-band significantly reduces the leakage current (from about 10 nA to about 2 nA) of edge pixels for all three crystals we tested with and without it. The energy resolution of the edge pixels of these Pt/Pt crystals is also improved from about 9 % to about 6.5 % with floating guard-band. Thus we find that the floating guard-band has significant effect for blocking contacts.

### 3.4 Grounded wall guard-band with Gold cathode and Gold anode

We also tested few crystals with Gold/Gold contacts mainly to verify the concept of the grounded guard-band. The results of our measurements are summarized in Figure 7c. We find that the Gold contacts exhibit blocking nature as expected[10]. The floating guard band is not effective for the one crystal we tested with and without it. We tested one crystal with the grounded guard-band deposited only on two side walls. The edge pixels for this crystal show quite poor performance. However, one crystal with a grounded guard-band on all walls, shows a promising result. The leakage current as well as energy resolution is uniform across all pixels. It can be seen that the leakage current is quite high for this crystal but this is the intrinsic property of the crystal. The same crystal with Indium contacts also show very high leakage current of about ~30 nA. It should be noted that these are results from just one crystal (though different crystal in each case) and hence should not be considered as a general trend. We hope to do a thorough study of the grounded wall guard band concept with a larger sample in future.

### 3.5 Hybrid contacts

We find that Indium contacts with IMARAD CZT give better uniformity across pixels of the same crystal however the energy resolution is poor. Platinum contacts, on the other hand, give better energy resolution but the uniformity is poor. Therefore to achieve the combination of both better uniformity and good energy resolution, we tried a few crystals with different metallic contacts on the anode and cathode surface. We tested two crystals with Platinum cathode, Indium anode and one crystal with Platinum cathode, Aluminum anode. Results of our measurements for these hybrid crystals are summarized in Figure 7d. We find that the Platinum/Aluminum crystal shows large variations between the leakage current of inner pixels and edge pixels. Similar variation is also seen in energy resolution among these pixels. The Platinum/Indium crystals, on the other hand, show much more uniform distribution of leakage current across all pixels. The edge pixels of Platinum/Indium crystals are still poor then inner pixels in terms of energy resolution. However, these results are from the measurements with only a few crystals. We need a larger sample of hybrid crystals to arrive at any concrete conclusion. Overall, crystals with hybrid contacts look promising.

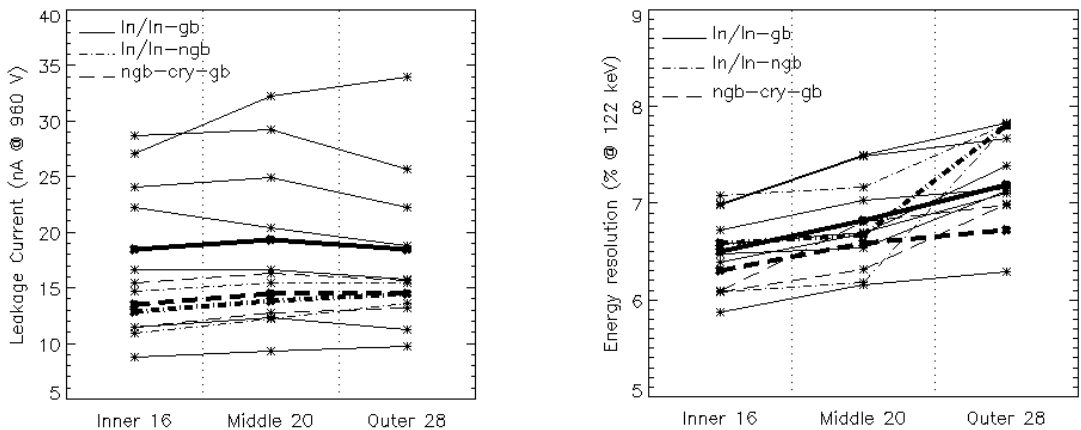

**Figure 7a:** Summary of leakage current and energy resolution for Indium/Indium contacts. Solid lines represent crystals with floating guard-band. Dash-dot-dash lines represent crystals without floating guard-band and long dash lines representing the same crystals with guard-band. Thin lines represent measurements of individual crystals whereas thick lines are averages of all crystals of respective types.

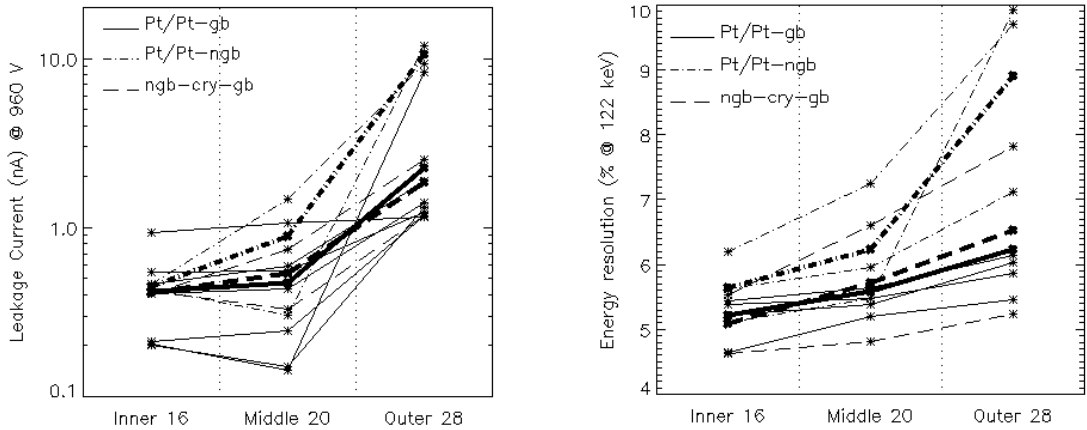

**Figure 7b:** Summary of leakage current and energy resolution for Platinum/Platinum contacts. The meaning of various line styles is same as Figure 7a.

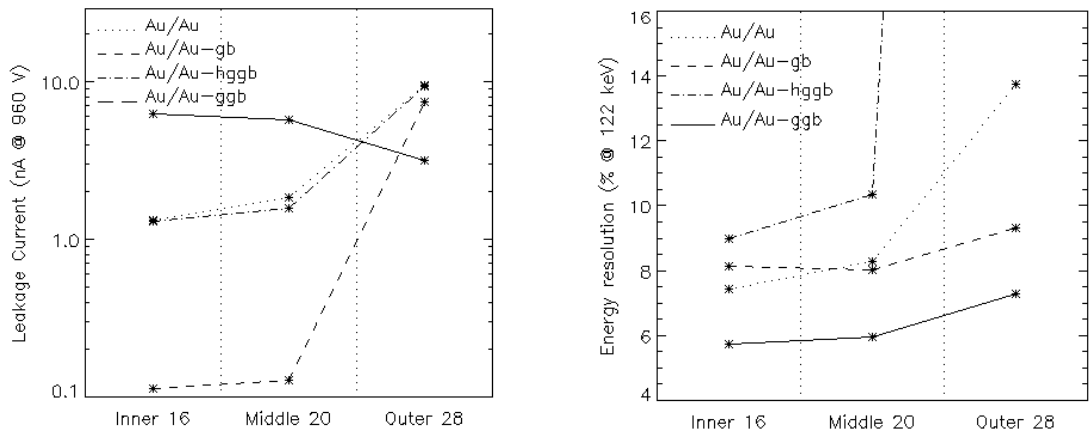

**Figure 7c:** Summary of leakage current and energy resolution for Gold/Gold contacts. Dotted and dashed lines represent one crystal with and without floating guard-band respectively. Dash-dot-dash and solid lines represent (different) crystals with grounded guard-band on two and four sides respectively.

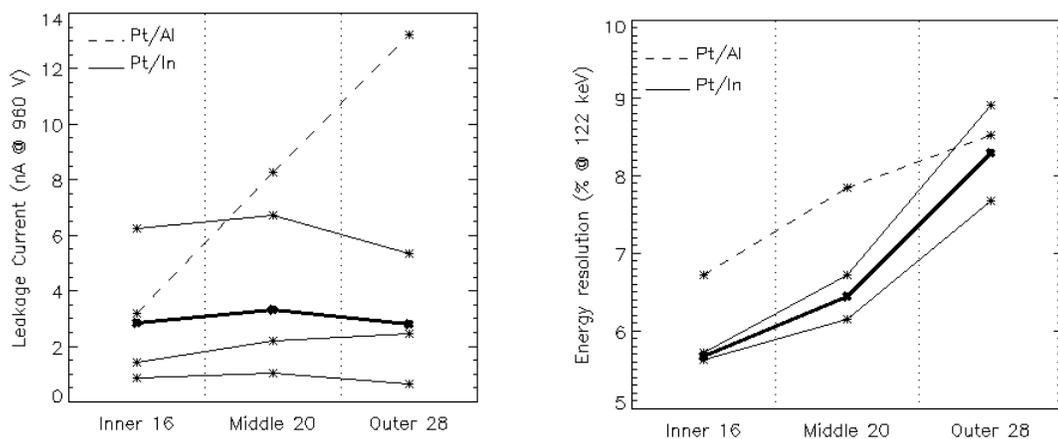

**Figure 7d:** Summary of leakage current and energy resolution for hybrid contacts. The dashed line represents one crystal with Platinum cathode and Aluminum anode. Solid lines represent crystals with Platinum cathode and Indium anode, with thin lines showing measurements of individual crystals, and thick line showing the average of all crystals of this type.

## 4. DISCUSSION

We carried out a comparative study of the full pixel behavior IMARAD CZT crystals with various metallic contacts. We find that ohmic contacts (i.e. Indium) show higher leakage current and larger energy resolution, compared to blocking contact (i.e. Platinum and Gold). With ohmic contacts we get 10 - 30 nA leakage current and 6 - 7 % energy resolution at 122 keV whereas with blocking contacts we get 0.1 - 1 nA leakage current and ~5 % energy resolution at the same energy. It should be noted that this result is poor compared to our previously reported measurements[10,11]. However these measurements used discrete front end pre-amplifier with very low noise. The XAIM 3.2 ASIC and the COB mounting assembly used in our measurements has intrinsic noise of about 600 $e^-$RMS. We are working on development of a very low noise 64 channel ASIC and we hope to achieve results similar to those with discrete pre-amplifier with the new ASIC. We believe that with low noise front-end electronics the difference between the performance of ohmic contacts and blocking contacts will increase with the blocking contacts providing much better performance.

However, ohmic contacts show better uniformity across the pixels of the same crystal. Platinum contacts, on the other hand, show systematically poor performance on the edge pixels, though we find that the edge pixel performance of blocking contacts can be improved by using the IMARAD floating guard band. The effectiveness of the floating guard band is surprising because it is just an adhesive tape with a thin metallic band deposited near one edge. The mechanism for the action of the floating guard band is not clear but it is thought that some capacitive effect is responsible for its effectiveness. In an attempt to improve the edge pixel performance with blocking contacts, We also tried a grounded guard band, which is basically an extension of the classical guard-ring. Initial results with one Gold/Gold crystal, for which we tried the grounded guard band, show considerable improvement in the performance of the edge pixels. However, we need more samples to arrive at any concrete conclusion.

The reason for the poor performance of the edge pixels with blocking contacts is not clear. However, one possible explanation can be that the contact resistance for blocking contacts is larger then the CZT bulk resistance whereas the contact resistance for ohmic contacts is smaller then the CZT bulk resistance. Then, since the contact resistance is larger then bulk CZT resistance for blocking contacts, electrons arriving at the anode surface find it easier to flow sideways rather then passing through the metallic contact. Finally when these electrons reach the edge, the only way to go is through the metallic contact, resulting in a much larger leakage current for the edge pixels. In the case of an ohmic contact, the contact resistance is smaller then the bulk CZT resistance and hence electrons arriving at anode surface anywhere in the crystal find it easier to pass through the contact resulting in a more uniform leakage current over all pixels. Regarding the radiation properties for edge pixels, we find that though leakage current and radiation properties such as energy resolution do not show a direct relation, they are not totally independent either. Thus larger leakage current for the edge pixels then leads to poor radiation performance on edge pixels. According to this model, the blocking contacts gives smaller leakage current whereas ohmic contacts give more uniform leakage current. This suggests that if a metal contact on the cathode surface has blocking nature it will result in low leakage current and if the metal contact on the anode has ohmic nature it will result in uniform leakage current. We carried out tests of crystals with hybrid metallic contacts to verify this and we find that our initial results for hybrid contacts (blocking cathode and ohmic anode) with IMARAD CZT crystals are consistent with this model, though testing with larger sample is required for confirmation. We are planning a detailed study with a larger sample of hybrid contact crystals in future.

## 5. CONCLUSIONS

We characterized large number of IMARAD CZT crystals with different metallic contacts. We find that there are systematic variations between the properties of inner and outer pixels. Such variations are much more prominent for the blocking contacts such as Platinum and Gold. We also find that the most of the observed variations between pixels are a property of the metallic contacts not the crystal itself because the same crystal with different contacts exhibits different types of variations. We find that for inner pixels, blocking contacts exhibit better performance then ohmic contact. We further tested the effect of IMARAD type floating guard-band as well as a new concept of a grounded wall guard-band. We find that both types of guard-band do improve properties of edge pixels with blocking contacts. We also characterized few crystals with blocking contact on cathode and ohmic contact on anode and find that this type of hybrid crystal provides promising performance. We plan to carry out further study with larger samples for the concepts of grounded wall guard-band as well as hybrid contact crystals.

## ACKNOWLEDGMENTS

This work is supported by NASA grants NAG5-5279 and NAG5-5396.